\documentclass[11pt]{article}

\usepackage{mathrsfs}
\usepackage{amsfonts}

\usepackage{amssymb,amsfonts}
\usepackage{graphicx}

\makeindex             

\begin{document}
\title{\textbf{Self-Organization In 1-d Swarm Dynamics}}
\author{Jiangning Chen, Weituo Zhang, Chjan C. Lim \\
Mathematical Sciences, RPI, Troy, NY 12180, USA\\
email: chenj27@rpi.edu} \maketitle

\begin{abstract}
Self-organization of a biologically motivated swarm into smaller
subgroups of different velocities is found by solving a
1-dimensional adaptive-velocity swarm, in which the velocity of an
agent is averaged over a finite local radius of influence. Using a
mean field model in phase space, we find a dependence of this
group-division phenomenon on the typical scales of the initial swarm in the position and velocity
dimensions. Comparisons are made to previous swarm models in which
the speed of an agent is either fixed or adjusted according to the
degree of direction consensus among its local neighbors.


\end{abstract}

\section{Introduction}

The dynamics of swarm behavior has long been a mystery in nature
\cite{12}-\cite{17}, and despite intensive work, remains an open
problem today \cite{1}-\cite{11}. Examples of swarms include flocks
of birds, schools of fishes and even crowds of people in cramped
public spaces. This research effort has introduced a wide range of
models for swarm behavior which is the background for our model in
this paper. In 1986, Reynolds \cite{1} simulated swarm behavior
according to the following rules: move in the same direction as the
neighbors; remain close to the neighbors; avoid collisions with
neighbors. In 1995, Vicsek et al. \cite{2} proposed an important
simpler model, in which all the agents have a constant speed and
only change their headings according to the other agents in the
influence radius (cf. also \cite{3,4,7,8}).

Other models for swarm behavior include \cite{5,6,8,9,10,11}. In
1995, Toner and Tu \cite{5} proposed a non-equilibrium continuous
dynamical model for the collective motion of large groups of
biological organisms. It describes a class of microscopic rules,
which includes the model in \cite{2} as a special case. In 2003,
Aldana and Huepe \cite{6} investigated the conditions that produce a
phase transition from an ordered to a disordered state in a model of
two-dimensional agents with a ferromagnetic-like interaction. In
their model, the agents still keep a constant speed and change their
headings as in \cite{2}. However, they are only influenced by their
neighbors in a fixed radius. Besides, Morale et al
\cite{9}-\cite{11} have proposed stochastic models for swarms from
2000. These models are based on a number of individuals subject to
several distinct mechanisms simultaneously - long range attraction,
and short range repulsion, in addition to a classical Brownian
random dispersal.

In 2008, Nabet, Benjamin, et al. \cite{7} established a model
simpler than \cite{2}, in which the agent speed is fixed and the
effects of agent position are ignored. The swarm in their model
includes two informed subgroups which have preferred directions of
motion and a third naive group that does not have a preference. They
investigate the tendency for members of the naive group to join the
other two groups and the conditions that govern this behavior. This
behavior to divide into subgroups is also what we focus on in our
paper. However, in our model, the division occurs spontaneously
without any special bias in the initial conditions of the swarm.

In 2011, W.Li and X.Wang \cite{8} proposed another model which is
also based on \cite{2}. In their model, each agent adjusts its
heading and speed simultaneously according to its local neighbors.
The change in speed at each time step depends only on the degree of
local direction consensus. Unlike our model, the new speed does not
depend on the previous speed or the speeds of nearby agents.

Motivated by the aim of a more complete understanding of swarm
behavior in the real world, we introduce a new model in
1-dimensional space, in which each agent continuously updates its
speed and direction according to the average velocity of the agents
within its influence radius. We simulate the model with initial
positions and velocities for the agents that are randomly drawn from
respective Gaussian distributions, and observe that there is a
robust non-equilibrium group-division phenomenon. We eventually
determine the time-scale of the group-division by analyzing the
rates of energy decrease in the system. It is our aim in this
1-dimensional model to isolate within a simple framework, one swarm
mechanism which is velocity-adaptive according to local velocity
averages. Despite its simplicity and spatial 1-dimensionality, the
robust group-division phenomenon that we found, appears to be
related to some aspects of the swarm behavior observed in the
natural world \cite{16,17}.

The paper is organized as follows. In section 2, we formulate the
discrete and mean field models in this paper. In section 3, we
discuss simulations conducted using these models to find the effects
of typical scales of initial position and velocity on the
group-division phenomenon. Then we examine the energetics of this
group-division phenomenon, and calculate the time when division
completes in section 4. Conclusions are given in section 5.

\section{models}
\subsection{Discrete model}

We consider a swarm of $N$ agents $\{a_1,a_2,...,a_N\}$ in 1-d space, each of which has velocity $v_i$ and position $p_i$, and assume that every agent can sense the other agents within the fixed radius $r$. We call these agents its neighbors and assume each agent adjusts its velocity according to the average velocity of the neighbors. For agent $a_i$, Let $n_i$ be the number of neighbors around it, and $\{a_{i1},a_{i2},...,a_{{in}_i}\}$ are corresponding neighbors, let $v_{ij}$ be the velocity of agent $a_{ij}$; according to the assumption, for each i = ${1,...,N}$:
$${dv_i \over dt}={1\over n_i}\sum_{j=1}^{n_i}v_{ij} -v_i.$$
We also have:
$${dp_i \over dt}=v_i.$$
Together, we get the equations in 2 dimensional phase space given by position and velocity.

Simulation of this model in two-dimensional phase space, as shown in Figure.1, reveals a group-division phenomenon in the dynamics. We measure the typical scales of initial position and velocity by the standard deviations ($\sigma(p)$, $\sigma(v)$) in the corresponding dimensions, define ($\sigma(p)$, $\sigma(v)$) as
$$\sigma(p)=\sqrt{{1\over {N-1}}\sum_{j=1}^{N}({p_j}-\overline{p})^2},$$
and
$$\sigma(v)=\sqrt{{1\over {N-1}}\sum_{j=1}^{N}({v_j}-\overline{v})^2}.$$
Then we generate the initial swarm using Gaussian distribution with typical scales $\sigma(p)$ and $\sigma(v)$ respectively. We run this model 20000 periods for 1000 agents, in each period, all the agents adjust their velocity in response to their neighbors. At the beginning, there is only one huge group; as the time goes by, the huge group starts to divide into several small groups, each group has its own velocity.

\begin{figure}[!hbtp]
\begin{center}
  \includegraphics[width=0.45\textwidth]{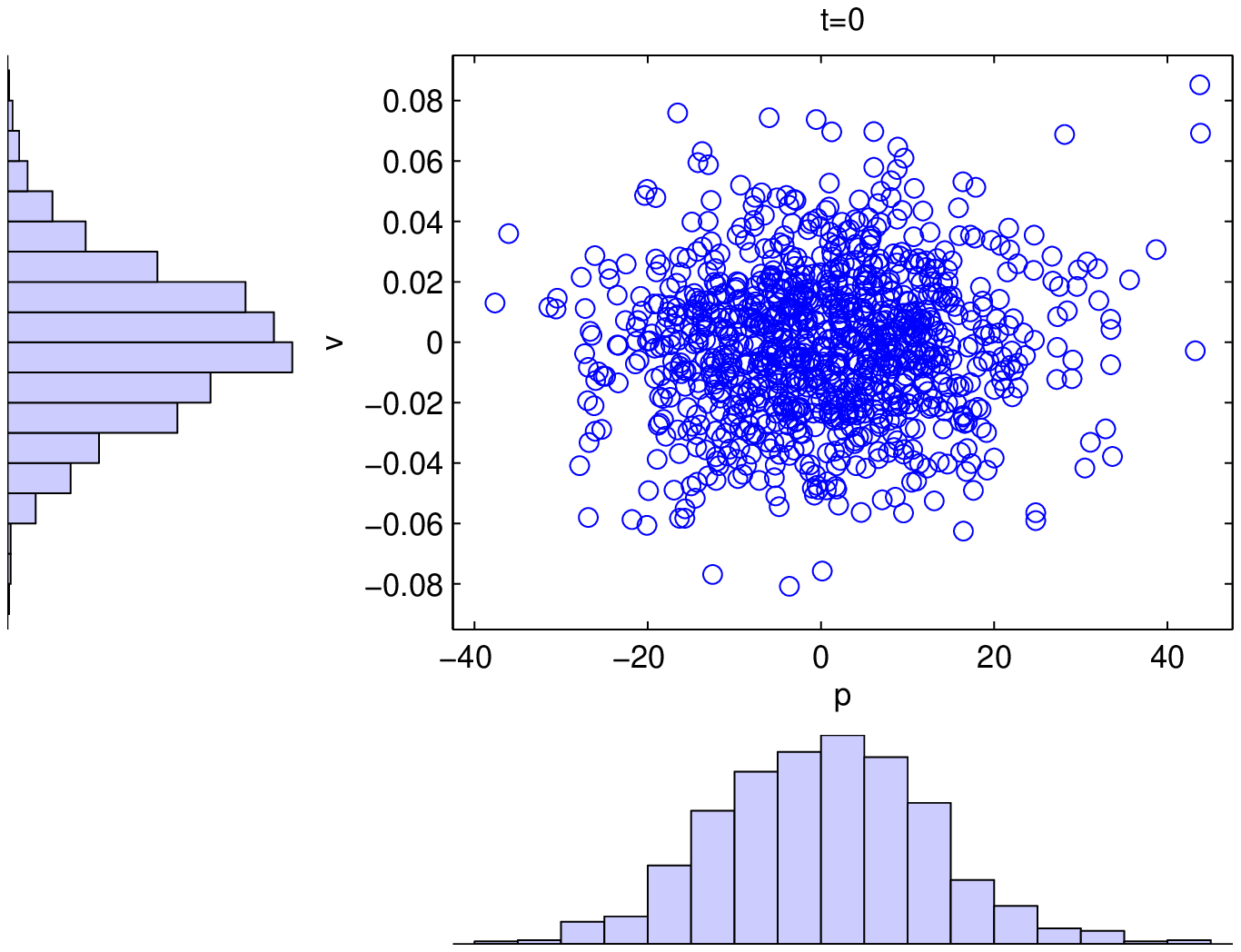} \includegraphics[width=0.45\textwidth]{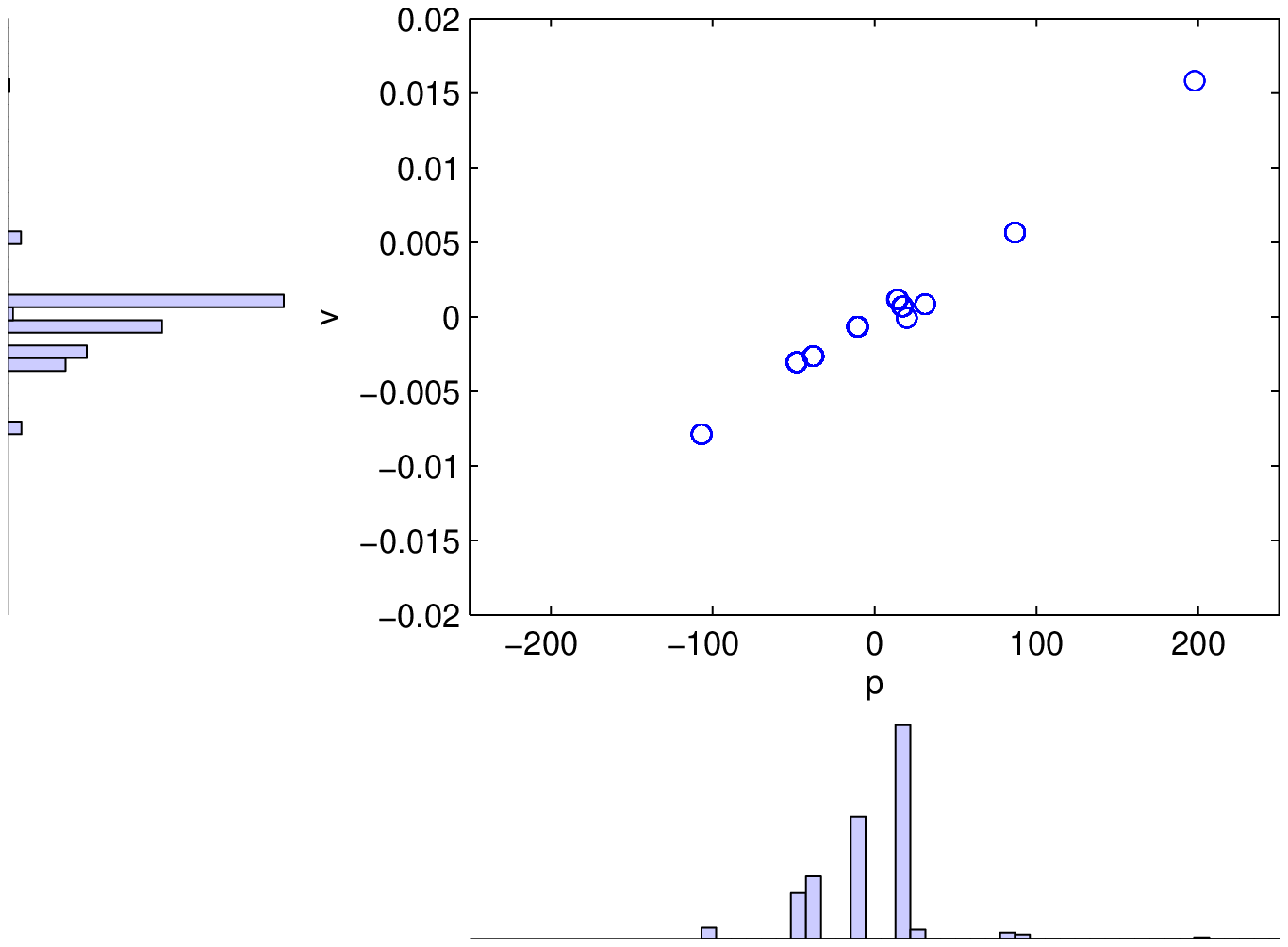}
  \caption{Initial condition: typical scale of $p$: $\sigma(p)=\sqrt{150}$; typical scale of $v$: $\sigma(v)=0.025$; influence radius: $r=0.05$; number of agents: $N=1000$; length of each time step: $dt=0.5$ and totally 20000 period. X-axis and y-axis represent position and velocity respectively. The histograms in the left and under the figure are the projection of the number of agents in the y-axis and the x-axis respectively. At the beginning of the simulation, there is only one group; as the time goes by, the group starts to divide into several small groups with most of the agents near the origin of the phase space.}
 \end{center}
\end{figure}

\subsection{Mean field model}

We use a mean field method to build a model in phase space. Assume the number of agents in the swarm is large enough; also the agents in any small area in phase space is well-distributed. Let $v,p$ be the velocity and position as before. Define $\rho$ as the probability measure in phase space --- for any point $x=(p,v)$ in phase space, $\rho(x)$ is defined as the probability for an agent to be in the neighborhood of $x$:
$$\rho(x)=\lim_{\varepsilon\rightarrow0}{prob\{a\in B(x,\varepsilon)\} \over \|B(x,\varepsilon)\|},$$
where $B(x,\varepsilon)$ denotes a ball of radius $\varepsilon$ centered at point $x$.

Let $t$ denote the time, $r$ denote the influencing radius. Define $u$ as the flow velocity in phase space, and define $Nb(p)=\{(p',v')|p'\in(p-r,p+r),v'\in(-\infty,\infty)\}$. Then
$$\int_{-\infty}^{\infty}(\int_{p_0-r}^{p_0+r}\rho(p,v)dp)\cdot dv=E[\mbox{number of agents in}\ Nb(p)],$$
where $E[\ ]$ denotes the expectation. Moreover:
$$\int_{-\infty}^{\infty}(\int_{p_0-r}^{p_0+r}\rho(p,v)\cdot vdp)\cdot dv=E[\mbox{total momentum in}\ Nb(p)].$$

Let $v_i$ be a random variable which denote the velocity of a agent in $Nb(p)$, $i=1,2,...,n$, where $n$ is the random variable which denote the number of neighbors. Since
\begin{eqnarray*}
E[\mbox{total momentum of agents in}\ Nb(p)]&=&E[\sum_{i=1}^{n}v_i]\\
                              &=&E[v_i]\cdot E[n]\\
                              &=&E[\mbox{average speed of agents in}\ Nb(p)]\cdot\\
                              & &E[\mbox{number of agents in}\ Nb(p)],
\end{eqnarray*}
thus
\begin{eqnarray*}
{\int_{-\infty}^{\infty}(\int_{p_0-r}^{p_0+r}\rho(p,v)\cdot vdp)\cdot dv \over \int_{-\infty}^{\infty}(\int_{p_0-r}^{p_0+r}\rho(p,v)dp)\cdot dv}
&=&{E[\mbox{total momentum of agents in}\ Nb(p)] \over E[\mbox{number of agents in}\ Nb(p)]}\\
&=&E[\mbox{average speed of agents in}\ Nb(p)]\\
&=&E[{1 \over n_i} \sum_{j=1}^{n_i}v_{ij}].
\end{eqnarray*}

Therefore, we have the mean field equation for the physical velocity:
$${\partial v \over \partial t}={\int_{-\infty}^{\infty}(\int_{p_0-r}^{p_0+r}\rho(p,v)\cdot vdp)\cdot dv \over \int_{-\infty}^{\infty}(\int_{p_0-r}^{p_0+r}\rho(p,v)dp)\cdot dv}-v,$$
and the phase space velocity:
$${\partial u(p_0,v_0) \over \partial t}=({\partial v \over \partial t},{\partial p \over \partial t})=({\int_{-\infty}^{\infty}(\int_{p_0-r}^{p_0+r}\rho(p,v)\cdot vdp)\cdot dv \over \int_{-\infty}^{\infty}(\int_{p_0-r}^{p_0+r}\rho(p,v)dp)\cdot dv}-v,v).$$

Finally, by conservation law in phase space:
\begin{equation}
{\partial \rho \over \partial t}=-\nabla \cdot (\rho u)=-{\partial \over \partial v}(\rho \cdot ({\int_{-\infty}^{\infty}(\int_{p_0-r}^{p_0+r}\rho(p,v)\cdot vdp)\cdot dv \over \int_{-\infty}^{\infty}(\int_{p_0-r}^{p_0+r}\rho(p,v)dp)\cdot dv}-v))-{\partial \over \partial p}(\rho v)
\label{equ1}
\end{equation}

Using the finite volume method \cite{18} under periodic boundary condition, we get the numerical solution of this integro-differential equation (\ref{equ1}). As shown in Figure.2, the numerical solution confirms the group-division phenomenon we find in the discrete model.

\begin{figure}[!hbtp]
\begin{center}
  \includegraphics[width=0.45\textwidth]{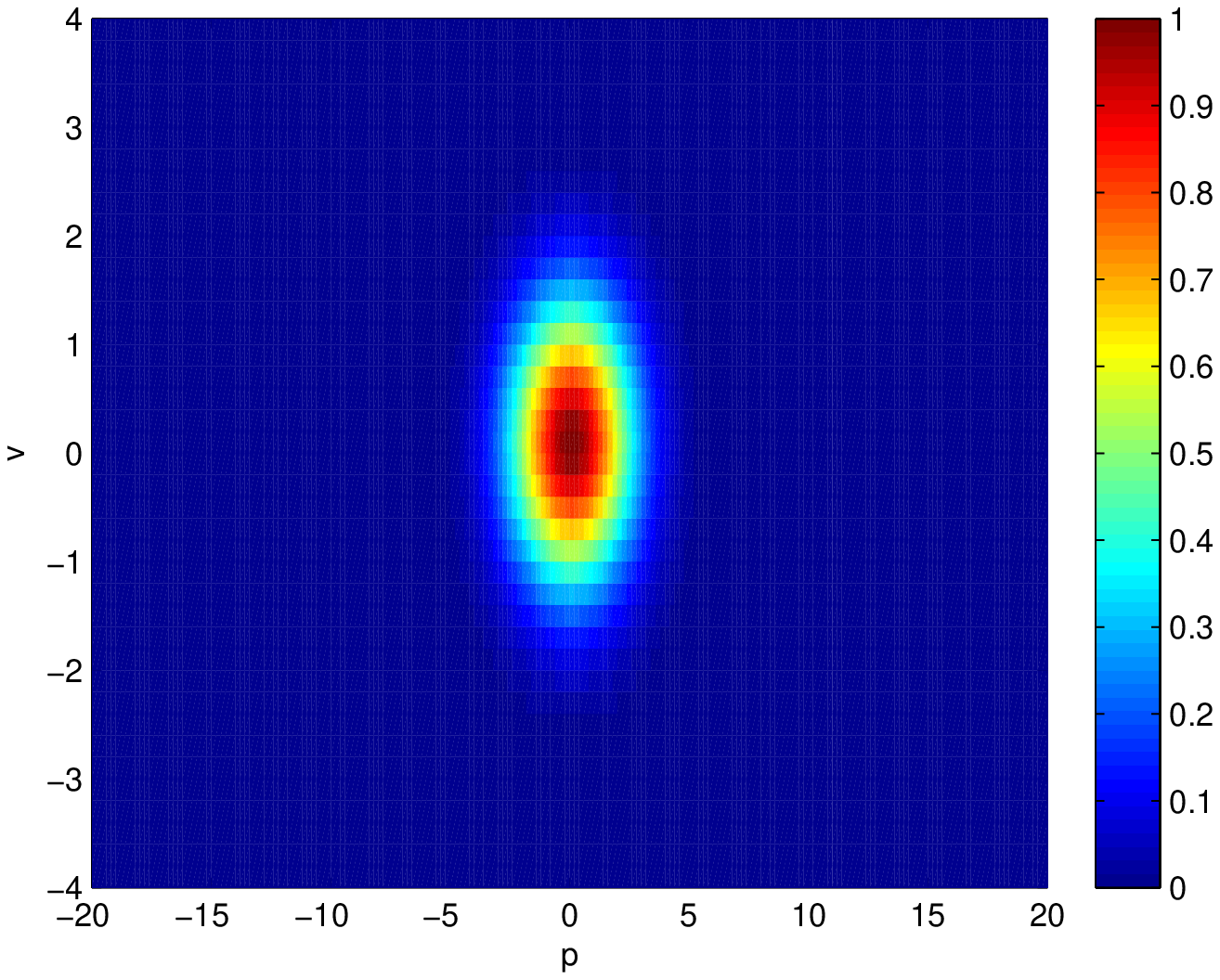} \includegraphics[width=0.45\textwidth]{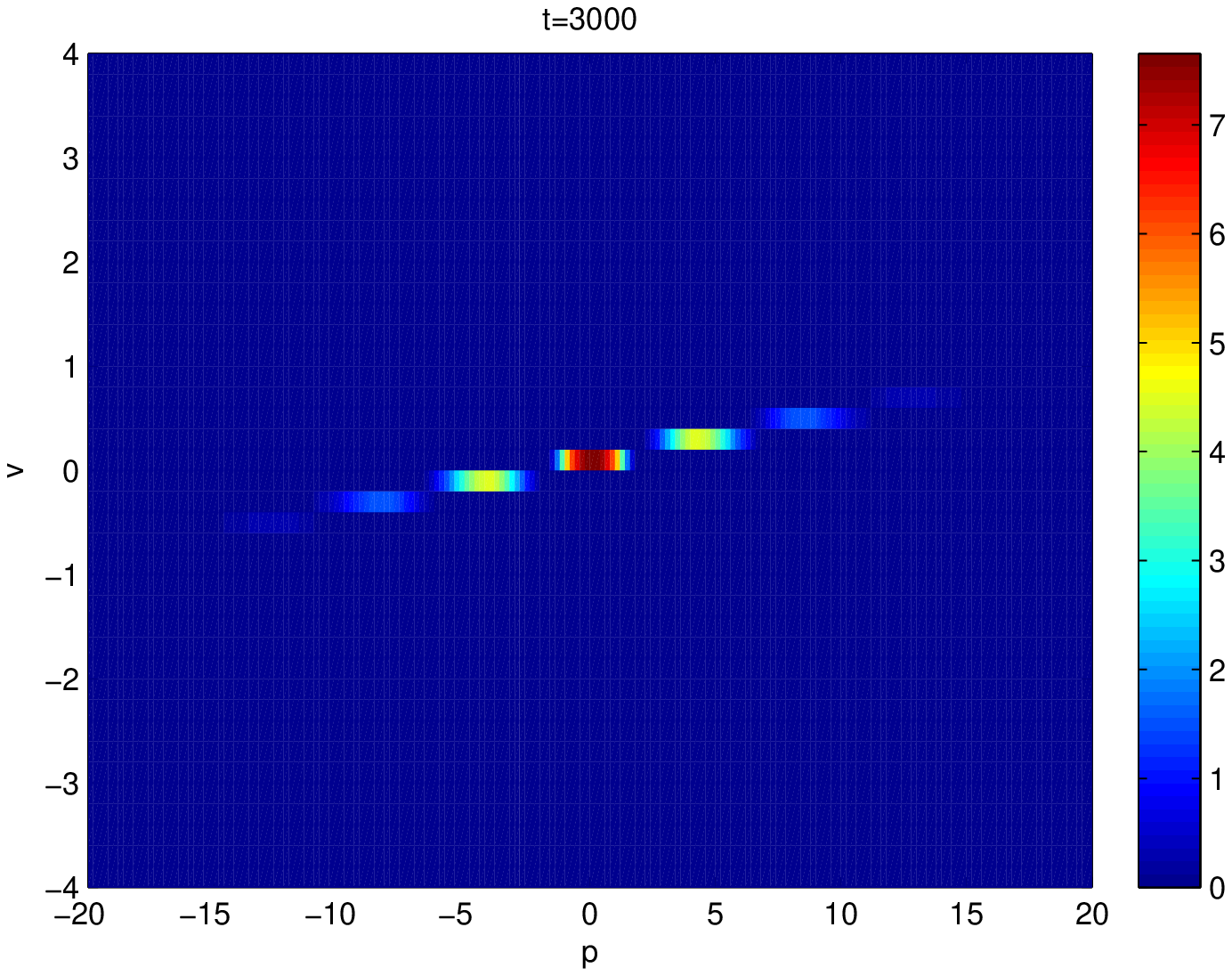}
  \caption{Initial condition: typical scale of $p$: $\sigma(p)=1.8$; typical scale of $v$: $\sigma(v)=0.9$; influence radius: $r=0.6$; length of each time step: $dt=0.005$ and totally 3000 period. x-axis is position and y-axis is velocity. The color reveals the density of the agents according to the right side color bar. At the beginning of the simulation, there is only one group; as the time goes by, the group starts to divide into several small groups. After 3000 periods, all agents have been divided into about 7 groups with most of the agents near the origin of the phase space.}
 \end{center}
\end{figure}

\section{Effects of typical scales of initial position and velocity on group-division phenomenon}

As shown above, the mean field model matches the discrete simulation very well. To see this more clearly, we display the data projection on the position axis. An example of such a projection, shown in Figure.3, reveals the group-division phenomenon in both models --- dash line and solid line represent mean field model and discrete model respectively. This comparison further validates the mean field model.

\begin{figure}[!hbtp]
\begin{center}
  \includegraphics[width=0.9\textwidth]{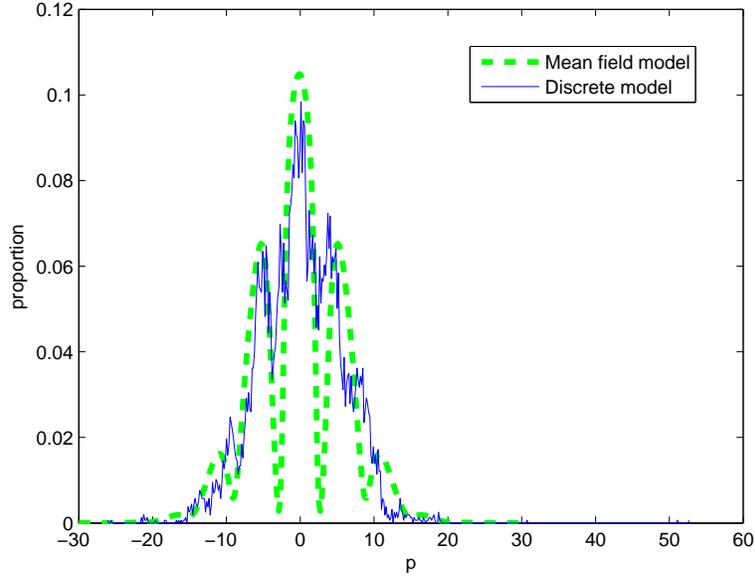}
  \caption{Initial condition: typical scale of $p$: $\sigma(p)=3$; typical scale of $v$: $\sigma(v)=1$; influence radius: $r=0.6$; length of each time step: $dt=0.005$; number of agents: $N=10000$; and totally 3000 period. x-axis is position and y-axis is the proportion of the agents in that position.}
 \end{center}
\end{figure}

Next, we find that the group-division phenomenon is related to the typical scales of initial position and velocity, as well as the influence radius. Scale $\sigma(p)$ and $\sigma(v)$ with $r$, we get some further examples, which are shown in the following (Figure.4). Y-axis indicates $\lambda$, the ratio of the size of largest group and the whole group. Smaller $\lambda$ causes more obvious group-division phenomenon, and the group-division phenomenon will disappear when $\lambda$ approaching 1. Since the line goes up all the way, it illustrates that as $\sigma(p)$ increases, most of the agent will stay in the largest group so that group-division phenomenon becomes weaker. Also, three different $\sigma(v)$ curves indicates that as $\sigma(v)$ increases, the group-division phenomenon become more pronounced.

\begin{figure}[!hbtp]
\begin{center}
  \includegraphics[width=0.9\textwidth]{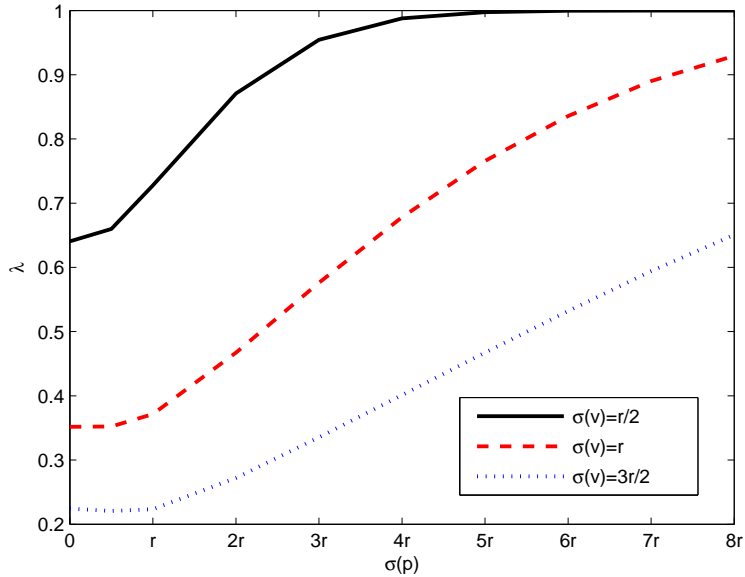}
  \caption{Initial condition: influence radius: $r=0.6$; length of each time step: $dt=0.005$; and totally 3000 period. x-axis is $\sigma(p)$, y-axis is $\lambda$, the ratio of the size of the largest group to the whole group.}
 \end{center}
\end{figure}

\section{Energy analysis}

In this section, we analyze energetics of the group-division
phenomenon. We find the energy-time curve has two stages connected
by a sharp transition(see figure.5).

Define $K$ as the total kinetic energy of the system. Under the group-division phenomenon,
label each group as:$\{g_1,g_2,...,g_n\}$, and the corresponding velocities of the
groups as:$\{v_{g_1},v_{g_2},...,v_{g_n}\}$. $N_{g_k}$ defines the number of agents in group $g_k$. Thus:
\begin{eqnarray}
K&=&\sum_{i=1}^{N}{v_i}^2 \nonumber\\
 &=&\sum_{i=1}^{N}(({v_i}-\overline{v})+\overline{v})^2 \nonumber\\
 &=&\sum_{i=1}^{N}({v_i}-\overline{v})^2+N\overline{v}^2 \nonumber\\
 &=&\sum_{k=1}^{N}\sum_{i\in g_k}[({v_i}-{v_{g_k}})+({v_{g_k}}-\overline{v})]^2+N\overline{v}^2 \nonumber\\
 &=&\sum_{k=1}^{N}\sum_{i\in g_k}[({v_i}-{v_{g_k}})^2+({v_{g_k}}-\overline{v})^2]+N\overline{v}^2 \nonumber\\
 &=&\sum_{k=1}^{N}\sum_{i\in g_k}({v_i}-{v_{g_k}})^2+\sum_{k=1}^{N}N_{g_k}({v_{g_k}}-\overline{v})^2+N\overline{v}^2.
\label{equ2}
\end{eqnarray}

According to the last line (\ref{equ2}), we write the
kinetic energy into three terms, where $\sum_{k=1}^{N}\sum_{i\in g_k}({v_i}-{v_{g_k}})^2$
is the first term, $\sum_{k=1}^{N}N_{g_k}({v_{g_k}}-\overline{v})^2$ is the second term, and $N\overline{v}^2$ is
the third term.

The first term is the relative kinetic energy of the agents with
respect to the center of mass of corresponding groups; the second
term is the relative kinetic energy of the groups with respect to
the global center of mass; the third term is the kinetic energy of
the global center of mass, which is almost constant from the
symmetry of Figure.3.

At the beginning of the simulation, since we only have one large
group:
$$\sum_{k=1}^{N}N_{g_k}({v_{g_k}}-\overline{v})^2=0.$$
At the end of the simulation, since we have several groups, and each agent
in its group will have the same velocity (i.e.:${v_i}={v_{g_k}}$):
$$\sum_{k=1}^{N}\sum_{i\in g_k}({v_i}-{v_{g_k}})^2=0.$$
Therefore, the kinetic energy shifts from the first term to the
second term, as it decreases monotonically from its initial value.

Moreover, since:
$${d({v_i}-{v_{g_k}})\over dt}=-(v_i-{v_{g_k}}),$$
solving this ODE yields:
$${v_i}={v_{g_k}}+{v_0}e^{-t}.$$
Therefore the first term could be written as:
\begin{equation}
\sum_{k=1}^{n}\sum_{i\in{g_k}}({v_i}-{v_{g_k}})^2=(\sum_{k=1}^{n}\sum_{i\in{g_k}}{v_{i_0}}^2)\cdot e^{-2t}.
\label{equ3}
\end{equation}

We consider the change of the variance of agent velocity, which can
be interpreted as kinetic energy. By equation (\ref{equ3}), we know
that the kinetic energy decrease exponentially in the first stage,
which is dominated by the first term of the equation. By taking the
log of the variance, the slope of the corresponding curve is found
to be close to -2. On the other hand, since we know that $v_{g_k}$
will finally be constant, the curve of kinetic energy should be a
straight line after division is complete, so the log of the variance
is still a straight line after division. Since we know that the
kinetic energy shifts from the first term to the second term, we
pick the point of the largest change in the semi-log variance during
the process as the end of the group-division phenomenon.

During the process, the variance of velocity decreases exponentially (see Figure.5).
We take the log of the variance of $v$(Y-axis in left figure), and consider
its rate of change of slope. The maximum point of change of slope rate,
which we take to be the end of division, is 195 (258) of the mean field model (discrete model).
The difference of these two numbers mostly come from the random choice of initial data in the discrete model,
so we consider the former one (195) is more accurate.

\begin{figure}[!hbtp]
\begin{center}
  \includegraphics[width=0.45\textwidth]{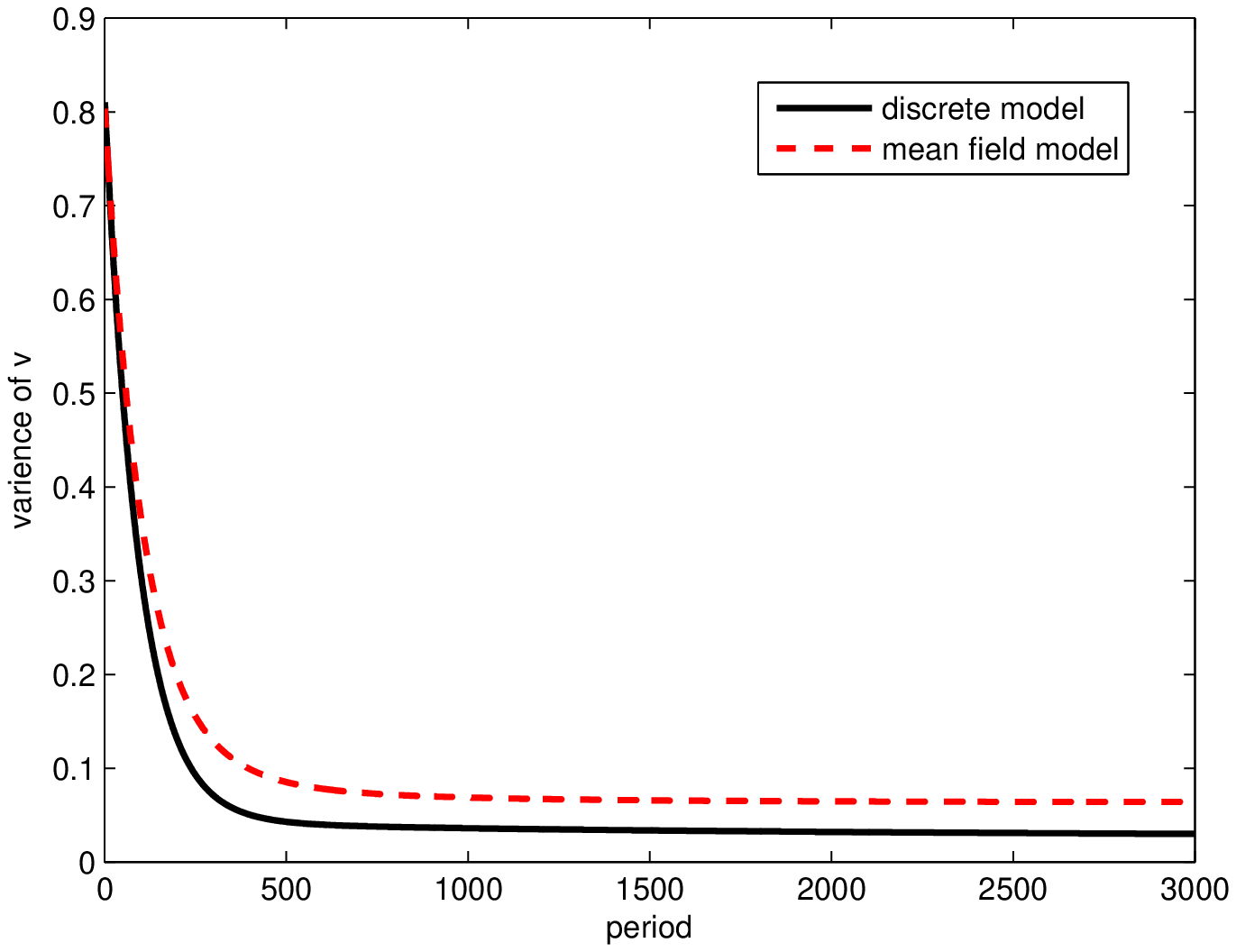} \includegraphics[width=0.45\textwidth]{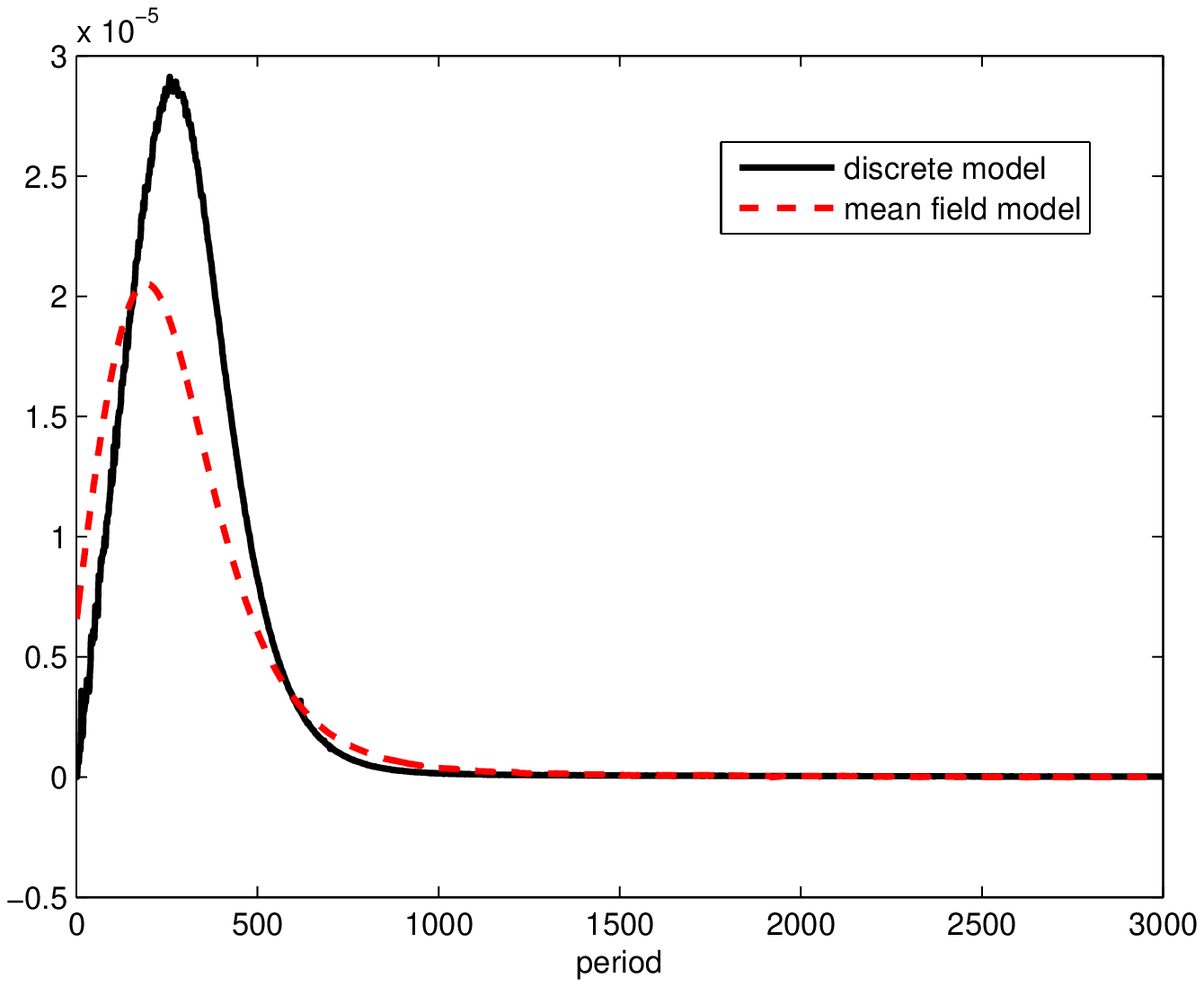}
  \caption{Initial condition: typical scale of $p$: $\sigma(p)=1.8$; typical scale of $v$: $\sigma(v)=0.9$; influence radius: $r=0.6$; length of each time step: $dt=0.5$ and totally 20000 period. For the left figure, X-axis shows the number of periods, Y-axis is the variance of velocity. The second derivation of the log of variance of velocity(Y-axis) in left figure gives the curve in the right figure.}
 \end{center}
\end{figure}

\section{Conclusion}

In our paper, we propose an adaptive-velocity swarm model in which
each agent not only adjusts its movement direction but also adjusts
its speed as a function of its local neighbors in 1-dimensional
space. Such a spatially 1-dimensional model is nonetheless capable
of supporting a robust group-division phenomenon that moreover,
depends strongly on just two key parameters of the initial swarm -
its typical scales in the initial positions and velocities of the agents. Details of possible weak
dependence on other properties of the initial swarm will emerge from
future studies. Furthermore, we show that group-division typically
occurs in two stages separated by the point in time whereby most of
the kinetic energy has shifted from the first term to the second
term. In the first stage, the whole group self-divides into several
smaller groups, followed by the second stage in which each smaller
group consolidates into one that has its own nearly constant
velocity and direction, within the fixed radius of influence.

Some difficult but important problems of our model remain to be
further investigated. For example, under what condition can we infer
the same result in 2-dimensional and 3-dimensional space? What kind
of behavior will emerge if certain agents have greater influence
than others? In the more applied aspect, how do we control the
motion of a swarm efficiently in our model? In addition, practical
stability analysis of the mean field model discussed in this paper
needs to be carried out.

\section*{Acknowledgements}
This work was supported in part by the
Army Research Office Grants No. W911NF-09-1-0254 and
W911NF-12-1-0546. The views and conclusions contained in this
document are those of the authors and should not be interpreted as
representing the official policies, either expressed or implied, of
the Army Research Office or the U.S. Government.


\end{document}